\documentclass[10pt,conference,letterpaper]{IEEEtran}
\IEEEoverridecommandlockouts
\usepackage{cite}
\usepackage{amsmath,amssymb,amsfonts}
\usepackage{algorithmic}
\usepackage{graphicx}
\usepackage{textcomp}
\usepackage{xcolor}
\def\BibTeX{{\rm B\kern-.05em{\sc i\kern-.025em b}\kern-.08em
    T\kern-.1667em\lower.7ex\hbox{E}\kern-.125emX}}
\def\*#1{\mathbf{#1}}

\begin{document}

\title{Non-Orthogonal Contention-Based Access for URLLC Devices with Frequency Diversity}


\author{\IEEEauthorblockN{Christopher Boyd\IEEEauthorrefmark{1}, Rados\l{}aw Kotaba\IEEEauthorrefmark{3}\IEEEauthorrefmark{2}, Olav Tirkkonen\IEEEauthorrefmark{1} and Petar Popovski\IEEEauthorrefmark{2}}
\IEEEauthorblockA{\IEEEauthorrefmark{1}Department of Communications and Networking, Aalto University, Finland\\
Email: \{christopher.boyd, olav.tirkkonen\}@aalto.fi}

\IEEEauthorblockA{\IEEEauthorrefmark{2}Department of Electronic Systems, Aalborg University, Denmark\\
\{rak, petarp\}@es.aau.dk}

\IEEEauthorblockA{\IEEEauthorrefmark{3}Intel Mobile Communications, Denmark\\}
}

\maketitle

\begin{abstract}
We study coded multichannel random access schemes for ultra-reliable low-latency uplink transmissions. We concentrate on non-orthogonal access in the frequency domain, where users transmit over multiple orthogonal subchannels and inter-user collisions limit the available diversity. Two different models for contention-based random access over Rayleigh fading resources are investigated. First, a collision model is considered, in which the packet is replicated onto $K$ available resources, $K' \leq K$ of which are received without collision, and treated as diversity branches by a maximum-ratio combining (MRC) receiver. The resulting diversity degree $K'$ depends on the arrival process and coding strategy. In the second model, the slots subject to collisions are also used for MRC, such that the number of diversity branches $K$ is constant, but the resulting combined signal is affected by multiple access interference. In both models, the performance of random and deterministic repetition coding is compared. The results show that the deterministic coding approach can lead to a significantly superior performance when the arrival rate of the intermittent URLLC transmissions is low. 
\end{abstract}

\begin{IEEEkeywords}
URLLC, grant-free, coded random access, MRC
\end{IEEEkeywords}

\section{Introduction}

Machine-Type Communication (MTC) is one of the main technologies in 5th Generation (5G) mobile communication. Within this very general category, we can distinguish two main use cases~\cite{5g} with widely differing requirements---massive MTC (mMTC), capable of supporting a large number of sporadically communicating devices, possibly battery-operated, and Ultra-Reliable Low Latency Communication (URLLC) enabling mission-critical MTC. Of the two, the latter especially has been igniting researchers' imaginations, as it would enable the implementation applications previously unattainable and considered futuristic, such as self-driving cars, remote surgery and telemetry, and more~\cite{5gusecase}.

While downlink communications in a cellular setting is fairly flexible, a radical change in the uplink access protocol might be necessary in order to fulfill the requirements of the more demanding MTC use cases. One solution is communication based on random access. For mMTC, this is motivated by the sporadic, infrequent traffic patterns which require energy efficient protocols, and the fact that the control overhead involved in establishing the connection significantly exceeds the amount of actual data to be transmitted. In URLLC, traffic also contains elements of randomness and is characterised by intermittent activation, but the random access is a means of achieving low latency levels, which could not be possible with the scheduling request/grant procedure in place. However, a solution based on random access is inherently unreliable, as it is subject to interference and collisions, so the random access for URLLC needs to be  augmented by other mechanisms that introduce redundancy to compensate for unavoidable collisions.

Multiple technologies to improve reliability of random access have been recently studied. Coded random access~\cite{paolini2015,choi2017} improves throughput and reliability by exploiting repetition coding and interference cancellation. Diversity slotted ALOHA has been considered by 3GPP as a potential solution for grant-free access~\cite{singh2018}, while in~\cite{IJWIN,globecomm}, the possibility to increase access reliability by preassigning non-orthogonal access sequences to users has been investigated.  In~\cite{kotaba2018}, a similar idea of preassigned patterns is treated, but with focus on the performance of successive interference cancellation (SIC) with imperfect channel state information (CSI). These technologies can be collectively called $K$-repetition schemes. 

In this paper we explore the diversity aspects of random access schemes based on packet repetition. If access opportunities that are exploited in a $K$-repetition scheme are independently fading, e.g., if access packets are transmitted over distinct frequency domain resource blocks experiencing different fading conditions, the reliability of communication is improved by diversity gains, in addition to the possible collision mitigation benefits. However, due to the fact that devices access the medium in a random manner, possibly causing collisions, this leads to a communication channel where {\it the diversity degree is a random variable}, governed by the arrival process of other users. We treat such a communication model by analyzing the probability distribution of the diversity degree, distribution of the contention level (number of simultaneous interferers) and total outage probability. We compare the performance of a simple receiver utilising a destructive collision-model, which discards all overlapping packets, with a more advanced system capable of optimally combining the replicas based on their signal-to-interference-plus-noise-ratio (SINR). Furthermore, we provide such analysis for the two coding approaches: uncoordinated, random selection of subchannels and deterministic assignment of patterns to the users. The latter technique is based on a code construction given by a Steiner system, as in~\cite{peeters2009,globecomm}.


\section{System Model}\label{Sec2}

We consider a communication system where $N$ URLLC users attempt to randomly access the uplink resources of a centralized receiver. Users active during a single timeslot transmit in an uncoordinated, grant-free fashion, and employ $K$-repetition coding of their access packets over $M$ orthogonal frequency subchannels. Users are slot synchronised to the receiver, and access packets are of equivalent size and occupy an entire subchannel. The users are assumed to become active randomly, such that the number of users active during a time instance follows a Poisson process with intensity $\lambda$.

Repetition coding of access packets provides robustness to inter-user collisions and facilitates diversity gain, which are integral for reliability in contention-based access over fading channels. Here we consider two approaches to coding: (i) users transmit $K$ packet replicas randomly over the slotted frequency resources, as in ALOHA-type schemes, and (ii) users transmit their replicas according to a deterministic and uniquely preallocated pattern from a designed access code. 

\begin{figure}[t!]	
	\centering
	\includegraphics[width=0.8\linewidth]{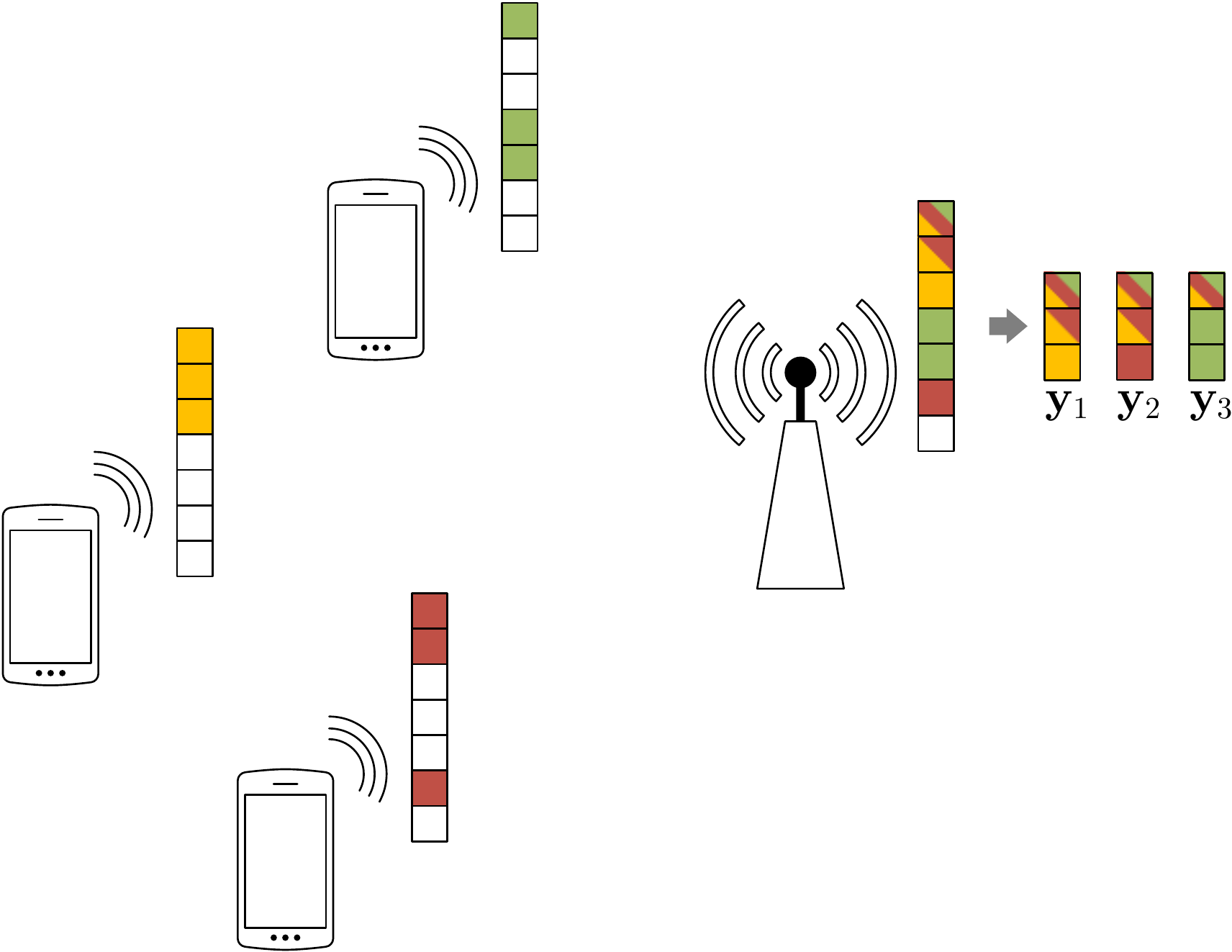}
	\caption{An example of the system with $N=3$ active users, $M=7$ frequency subchannels and $K=3$ repetitions of each packet. Users $1$ and $2$ manage to have a single interference-free replica, while the $3^{rd}$ user has two. The subchannels where collisions occur might be used to further increase the reliability if a more advanced processing is available.}
	\label{fig:1}
\end{figure}

Two receiver models are investigated: (i) a destructive collision model, a PHY layer abstraction to the MAC layer which assumes that colliding packets are lost and only interference-free packets may be correctly received, and (ii) a multi-user interference (MUI) model, where all packets are used to decode the signal but their contribution depends on their effective SINRs. In this paper, maximum-ratio combining is applied to the different packet replicas to achieve this.

The received complex baseband signal corresponding to the symbol $x_j$ transmitted by a device $j$ reads
\begin{equation}
	\*{y}_j=\*{h}_j x_j + \sum_{l=1}^{L_j}\*{g}_{j,l} z_{j,l} + \*{n}_j = \*{h}_j x_j + \*{i}_j,
\end{equation}
where $L_j$ is a random number of interferers perceived by user $j$, $\*{h}_j, \*g_{j,l} \in \mathbb{C}^K$ are the channel gains of the signal of interest and its $l$-th interferer respectively (which are assumed to be known at the receiver), $z_{j,l} \in \mathbb{C}$ are the interfering symbols,  $\*{n}_j \in \mathbb{C}^K$ is the additive white Gaussian noise (AWGN) with zero mean and variance $N_0$, and $\*{i}_j \in \mathbb{C}^K$ is the joint interference-plus-noise term. 
Note, that interferers might occupy only some of the slots of $j$, in which case the remaining entries of $\*{g}_{j,l}$ are $0$. We define the linear filter
\begin{equation}
	\*{f}_j=\*{W}_j \*{h}_j \label{eq:weights}
\end{equation}	
where $\*{W}_j=\mathrm{diag}[w_{j,1},w_{j,2},..,w_{j,K}]$ is a diagonal matrix of real-valued reliability weights that depend on the combining strategy. Applying the combiner yields
\begin{equation}
	r_j=\*{f}_j^H \*{y}_j.
\end{equation}
The post combining SINR of $j$-th user's signal is approximated in uncorrelated interference by
\begin{eqnarray}
	\gamma_j &=& \frac{\left|\*{h}_j^H \*{W}_j \*{h}_j\right|^2}{\mathrm{E}\left\{ \left| \*{h}_j^H \*{W}_j \*{i}_j \right|^2 \right\}} \nonumber\\ 
	&=& \frac{\left(\sum_{k=1}^K w_{j,k} \left| h_{j,k} \right|^2 \right)^2}{\sum_{k=1}^K w_{j,k}^2 \left| h_{j,k} \right|^2 \left(\sum_{l=1}^{L_j} \left| g_{j,l,k} \right|^2 + N_0 \right)}~,\label{eq:5}
\end{eqnarray}
where the reliability weights for MRC in the multi-user interference model that maximize $\gamma_j$ are given by
\begin{equation}
	w_{j,k}=\frac{1}{\sum_{l=1}^{L_j} |g_{j,l,k}|^2 + N_0}~.\label{eq:6}
\end{equation}

In the destructive collision model only those $0 \leq K' \leq K$ replicas which were received collision-free can be combined together. Let us further denote by $\mathcal{I}_j$ a subset of indices which correspond to those packets. Then, the signal model can be simplified since $\*i_j=\*n_j$ and
\begin{equation}
	w_{j,k} = 
	\begin{cases}
		1 & \text{for } k \in \mathcal{I}_j \\
		0 & \text{otherwise}
	\end{cases},
\end{equation}
resulting in the final signal-to-noise-ratio (SNR)
\begin{equation}
	\gamma_j = \sum_{k \in \mathcal{I}_j} \frac{|h_{j,k}|^2}{N_0}~.
\end{equation}
In the remainder of this paper we will often discuss a signal from the perspective of a single device and omit the index $j$ whenever it does not create ambiguity.


\section{Available Diversity after Collisions}\label{Sec3}

Consider a timeslot in which a given user $U$ of population $N\geq 1$ is active and transmitting $K$ packet replicas randomly over the $M$ access resources, along with $N-1\sim \mathrm{Poisson}(\lambda)$ simultaneously active users. The probability that $K'$ of $K$ replicas are received without interference depends on the coding strategy. 

\subsection{Diversity Slotted ALOHA}


In diversity slotted ALOHA (DSA), users transmit their $K$ packet replicas over the $M$ subchannels randomly, following a uniform distribution. From the perspective of user $U$, the probability that the $N-1$ other users collide is such a way that $K'$ of $K$ subchannels are unoccupied by packet replicas follows from the classical occupation problem, and is given by
\begin{equation}
p_r(K'|N) = \binom{K}{K_{\mathrm{diff}}} \sum_{n=0}^{K_{\mathrm{diff}}} (-1)^n a_n X_n^{N-1}~, \label{eq:1}
\end{equation}
where $K_{\mathrm{diff}}=K-K'$, $a_n = \binom{K-K'}{n}$, $X_n = \binom{M-K'-n}{K}/\binom{M}{K}$, 
$p_r(K'\neq K|1)=0$, and $p_r(K|1)=1$.

The probability that user $U$ occupies $K'$ interference-free subchannels, conditioned on the arrival process, is
\begin{eqnarray}
\!\!\!\!\! p_r({K'}) \!\!&=& \!\!\!\! \sum_{N=1}^{\infty} p_r(K'|N) p(N-1)\nonumber \\
&=& \!\!\!\! \binom{K}{K_{\mathrm{diff}}} \sum_{N=1}^{\infty} \sum_{n=0}^{K_{\mathrm{diff}}} (-1)^n a_n \frac{(X_n\lambda)^{N-1}}{(N-1)!} e^{-\lambda}  \label{eq:2}
\end{eqnarray}

\subsection{Designed Codes }

Designed and uniquely preallocated user codes have been shown to outperform the random coding approach in a URLLC context~\cite{Paolini,IJWIN}. Such codes limit the number of supportable users in order to coordinate the interference over that population. The performance of combinatorial code designs such as Steiner system $S(t,K,M)$ as random access codes has been explored in~\cite{peeters2009,globecomm}. Here we consider Steiner $t=2$ designs, as their highly symmetric structure makes for ready analysis. Note that $t>2$ designs may produce significantly larger codes, and may therefore be more practical. 

Consider the scenario where each of the $N$ users is uniquely allocated a repetition pattern from a $S(2,K,M)$ code~\cite{lajolla}. The maximum supportable user population is limited to $C=|S(2,K,M)|=M(M-1)/K(K-1)$. However, the structure of the code ensures that the number of users that may overlap with user $U$ in a single subchannel is at most $D=(M-K)/(K-1)$. When user $U$ is active, the $N-1$ simultaneously active users will be employing repetition patterns from the remaining $C-1$, of which at most $kD$ can overlap user $U$ in $k$ slots. The probability that user $U$ has $K'$ of $K$ diversity branches post collisions is therefore
\begin{equation}
	p_{\mathrm{det}}(K'|N) = \binom{K}{K_{\mathrm{diff}}} \sum_{n=0}^{K_{\mathrm{diff}}} (-1)^n a_n Y_n~,
\end{equation}
where $Y_n = \binom{(C-1)-D(n + K')}{N-1}/\binom{C-1}{N-1}$, and the same restrictions on $N$ as in~(\ref{eq:1}) apply. The probability $p_{\mathrm{det}}({K'})$ can be found similarly to~(\ref{eq:2}), with $p_{\mathrm{det}}(K'|N)$ in place of 	$p_r(K'|N)$. 

\begin{figure}[t!]	
	\centering
	\includegraphics[width=80mm,trim={0 0 0 7.5mm},clip]{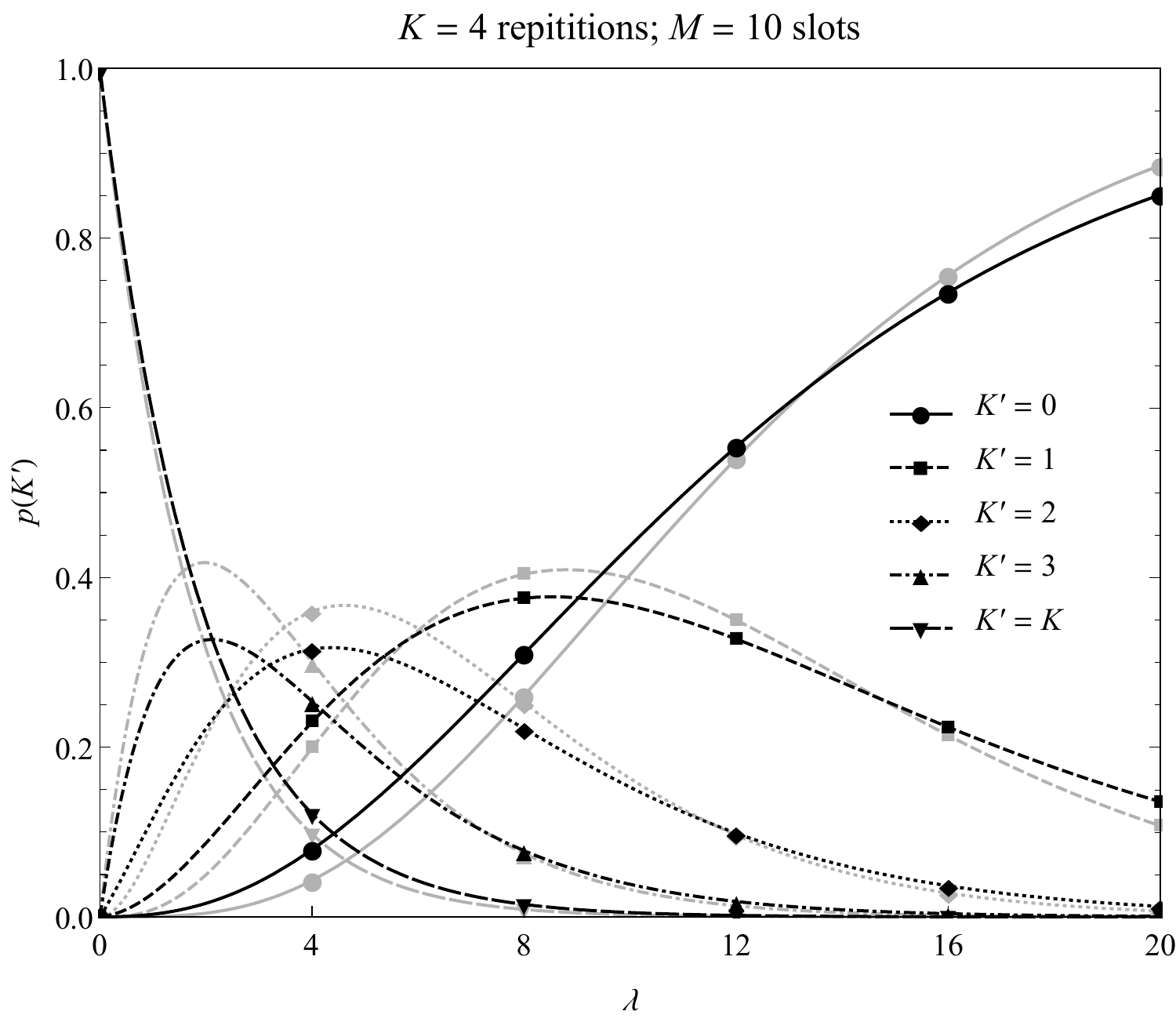}
	\caption{Probability distribution of the available diversity for DSA (black) and Steiner (grey) with $M=25$ and $K=4$ under Poisson arrivals.}
	\label{fig:2}
\end{figure}

Figure~\ref{fig:2} compares the probability of $K'$ diversity branches being available post collisions for the DSA scheme over $M=25$ subchannels with repetition factor $K=4$, and a deterministic coding scheme employing a $(2,4,25)$ Steiner system with $C=50$, as a function of the arrival intensity $\lambda$. Evident in the plot is how, at lower intensities, the Steiner code trades-off the probability of producing the best outcome ($K=K'$) to increase the probability of good outcomes (e.g. $K=3$), and reduce the probability of the worst outcome ($K'=0$). Since this worst outcome is especially detrimental in the collision model, we can expect significant deterministic gain in this intensity region. As $\lambda$ approaches $M$, the structure of the deterministic code becomes a disadvantage. If the URLLC users are activated in an intermittent and uncorrelated manner, then the expected number of users simultaneously active during a given slot is $\lambda<<M$. 


\section{Interferers per Subchannel}\label{Sec4}

Consider again a user $U$ transmitting during a timeslot with $N-1$ other users. In the case of weighted MRC, we are interested in the number of packets from $L$ interferers present in the subchannels occupied by $U$. Let $0\leq L'\leq L$ be the number of independently Rayleigh faded packets from the $N-1$ interfering users \emph{in a given subchannel} of user $U$. The probability distribution of $L'$ depends on the coding strategy. 

\subsection{Diversity Slotted ALOHA}

In DSA, the repetition coding procedure amounts to users independently selecting one of the $\binom{M}{K}$ possible binary patterns with replacement. As such, the maximum number of interferers observed by $U$ in one subchannel is $N-1$. The probability of $L'$ interfering packets in a given subchannel occupied by user $U$ is given by 
\begin{equation}
	p_r(L'|N) = \frac{ \binom{M-1}{K}^{N-1-L'}}{\binom{M}{K}^{N-1}}  \binom{M-1}{K-1}^{L'} \binom{N-1}{L'}~. \label{eq:3}
\end{equation}
Note that this is an approximation assuming the interferers select the subchannels independently. The probability $p_r(L')$ can be found as in~(\ref{eq:2}) by marginalizing over Poisson distributed $N$.

\subsection{Designed Codes}

With a finite set of $C$ deterministic access patterns, the probability that user $U$ sees $L'$ interferers in a given subchannel in which it is active, is the probability that $L'$ of the $N-1$ other users are using patterns from the $D$ that overlap in that subchannel. Since patterns are uniquely preallocated, selection from the $C-1$ available codes is done without replacement. The random variable $L'$ therefore follows the hypergeometric distribution, such that
\begin{equation}
	p_\mathrm{det}(L'|N) = \binom{C-1}{N-1}^{-1} \binom{D}{L'} \binom{C-1-D}{N-1-L'}~,  \label{eq:4}
\end{equation}
and $p_\mathrm{det}(L')$ is found as in~(\ref{eq:2}).

\begin{figure}[t!]	
	\centering
	\includegraphics[width=80mm,trim={4.65cm 0 4.5mm 7.5mm},clip]{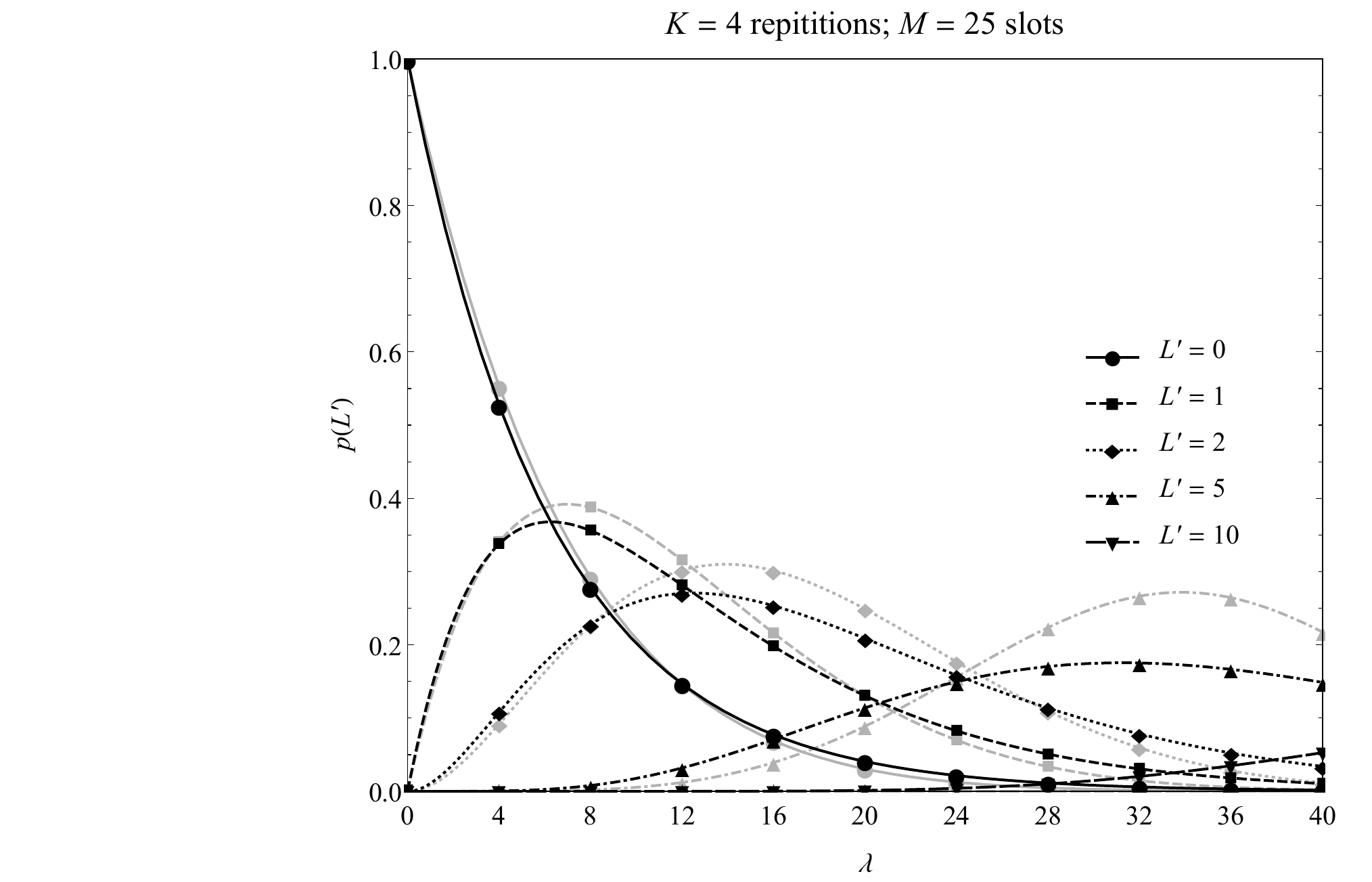}
	\caption{Probability distribution of the number of interferers in a subchannel for DSA (black) and Steiner (grey) with $M=25$ and $K=4$ under Poisson arrivals.}
	\label{fig:3}
\end{figure}

Figure~\ref{fig:3} compares the probability distributions of $L'$ for the DSA scheme over $M=25$ subchannels with $K=4$ and the $(2,4,25)$ Steiner code with $C=50$, as functions of $\lambda$. Here, the Steiner codes slightly increases the probability of the best outcome ($L=0$) at lower intensities, but decreases it as $\lambda$ approaches $M$. More pronounced is how the Steiner code increases the probability of lower numbers of interferers (e.g. $L=1$) in certain windows of intensity.


\section{Diversity Combining}\label{sec5}

Lastly, let us analyse the outage probability performance of the different schemes. The outage probability is defined as the probability that the post-processing SINR $\gamma$ falls below a certain threshold $\theta$, i.e
\begin{eqnarray}
    p_{\mathrm{out}} &=& p(\gamma<\theta)~.
\end{eqnarray}
This metric will depend on both the coding technique as well as the applied receiver processing.

\subsection{Collision Model}

After discarding the packets which experienced collision, the remaining $K'$ replicas transmitted by user $U$ can be combined by the receiver. Assuming perfect CSI is available, and the SNR of a single packet is exponentially distributed (following the Rayleigh fading assumption), the post processing SNR has the distribution
\begin{eqnarray}
	p(\gamma) &=& \sum_{K'=0}^{K}p(\gamma|K')p_c(K')\\	
	&=& \sum_{K'=0}^{K}\frac{1}{(K'-1)!} \frac{\gamma^{K'-1}}{\Gamma^{K'}} e^{-\gamma/\Gamma} p_c(K')~, \label{eq:7}
\end{eqnarray}
where $\Gamma$ is the expected SNR per packet and $p_c(K')$ is either $p_r(K')$ or $p_{det}(K')$ depending on the scenario. The \eqref{eq:7} follows from the fact that the sum of $K'$ exponentially distributed random variables with scale $\Gamma$ is $Gamma(K',\Gamma)$ distributed.


\subsection{Multi-User Interference Model}

\begin{figure}[t!]	
	\centering
	\includegraphics[width=80mm,trim={0 0 0 7mm},clip]{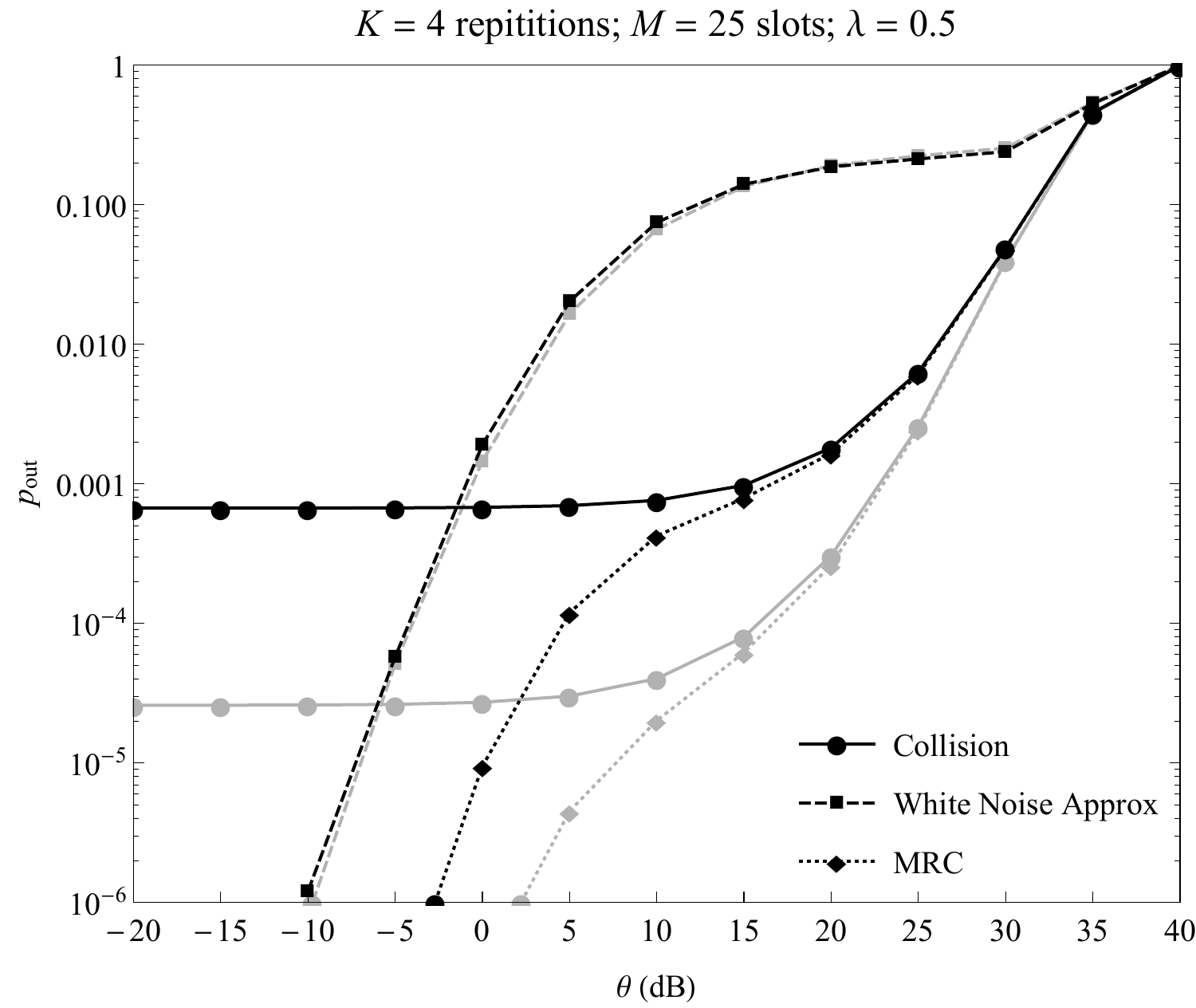}
	\caption{Probability of outage for DSA (black) and Steiner (grey) with $M=25$, $K=4$ and $\Gamma = 30$ dB, under Poisson arrivals with $\lambda = 0.5$.}
	\label{fig:4}
\end{figure}

\begin{figure}[t!]	
	\centering
	\includegraphics[width=80mm,trim={0 0 0 7mm},clip]{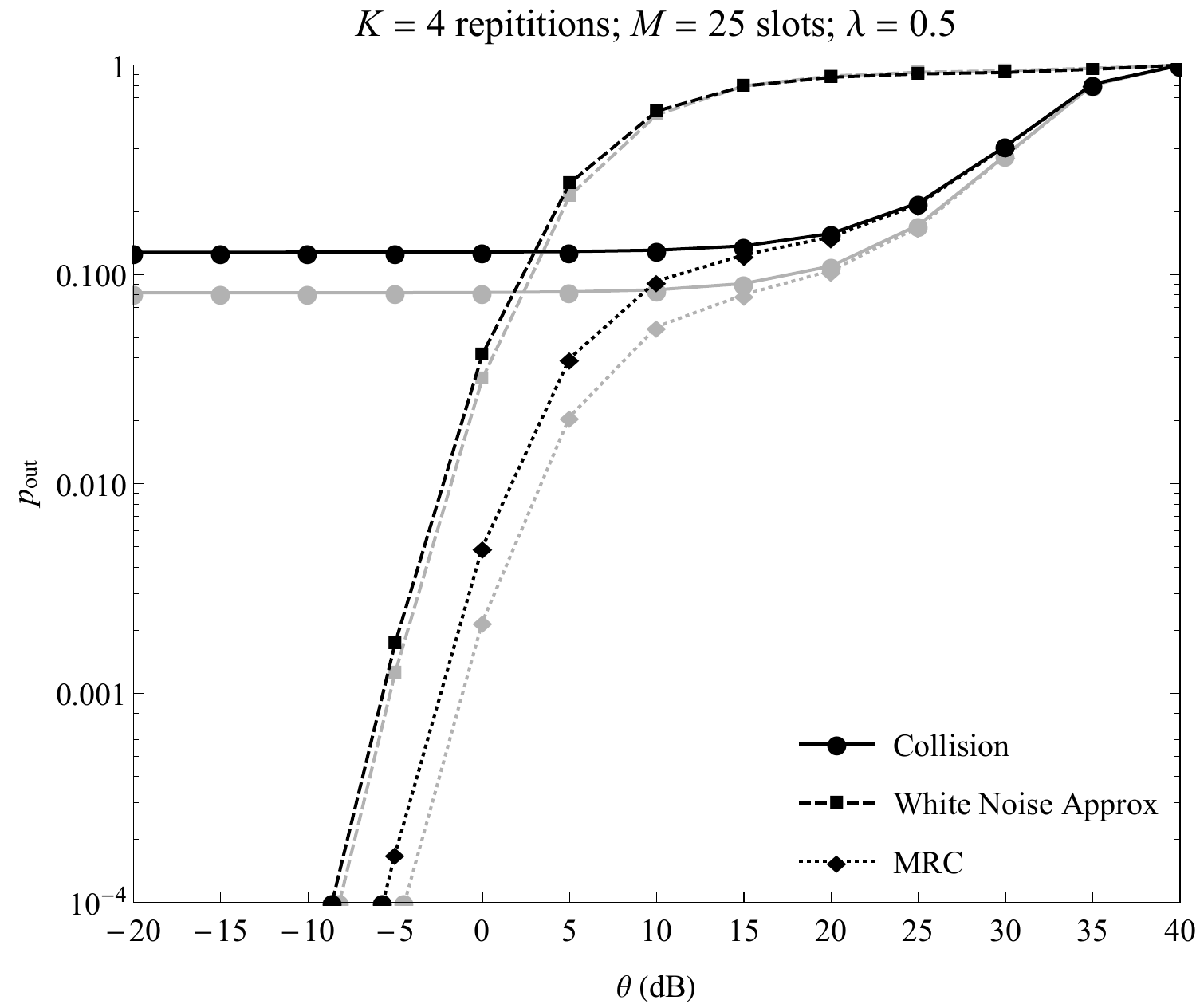}
	\caption{Probability of outage for DSA (black) and Steiner (grey) with $M=25$, $K=4$ and $\Gamma = 30$ dB, under Poisson arrivals with $\lambda = 5$.}
	\label{fig:5}
\end{figure}

In the case of MUI, obtaining closed form expressions of even the conditional SINR distribution is not feasible, as it quickly becomes intractable (i.e. for more than one interferer)
\begin{eqnarray}
	p(\gamma|L') = \int_{N_0\gamma}^{\infty} f_{\rm exp}\left(x|\Gamma \right) f_{\rm gamma}\left(\frac{x}{\gamma}-N_0|L',\Gamma \right) dx~.
\end{eqnarray}
Furthermore, the full distribution would require marginalizing the convolution of individual SINRs of the replicas over all $N$ and all possible realizations of $L'_1,...,L'_K$
\begin{equation}
\begin{aligned}
	p(\gamma) = \sum_{N=1}^{\infty} \sum_{L'_1}^{N-1} \dots \sum_{L'_K}^{N-1} & \left( p(\cdot | L'_1) \ast \dots \ast p(\cdot | L'_K) \right)(\gamma)\\
	& \times p\left(L'_1,\dots ,L'_K | N  \right)p(N)~.
\end{aligned}
\end{equation}
To obtain $p(\gamma)$, and eventually $p_{\mathrm{out}}$, for the multi-user interference model we resort to simulation. We generate multiple instances of $(N,\*H)$, i.e. number of transmitting devices, their channel gains and patterns accordingly (DSA or Steiner), and evaluate the effective SNR according to~\eqref{eq:5} and~\eqref{eq:6}. Figures~\ref{fig:4} and~\ref{fig:5} show the outage probability performance as function of the SINR threshold $\theta$, for the MRC receiver in the collision and MUI models, with random and deterministic repetition coding. Additionally, included in the plots is the outage probability for a white noise approximated match filter (WN-MF), for which the reliability weights in (\ref{eq:weights}) are set to $w_k=1$ for all packets.

For low access intensity, represented here by Figure~\ref{fig:4} with $\lambda=0.5$, the gains offered by deterministic codes are significant compared to the uncoordinated traffic, e.g. at $\theta=5$dB the difference in offered reliability is more than an order of magnitude (and further two orders of magnitude compared to the white noise approximation). This gain diminishes as the intensity of traffic increases (cf. Figure~\ref{fig:5}), since the structure of the Steiner code becomes irrelevant as the channel becomes flooded with packet replicas. The collision model exhibits a clear error floor related to $p(K' = 0)$.

With regard to required processing and complexity, the receiver in the collision model is the simplest---it only needs to detect whether or not there were collisions in the subchannels occupied by a given user, and measure the SNR of each of their $K'$ interference-free packets. The MRC receiver, however, requires accurate measurements of the channel gain (and phase), as well as a precise estimate of the interference and noise corrupting each packet. 
The white noise approximation requires precise channel information for each packet, but does not need an interference plus noise estimate. The white noise approximation shows to what degree neglecting interference plus noise whitening is detrimental to the performance of the MRC receiver. In collision channels of the type discussed here, noise plus interference whitening is of prime importance.


\section{Conclusion}

We have presented a study of multichannel random access mechanisms for supporting low-latency uplink transmission from a set of uncoordinated devices. This study is in the context of Ultra-Reliable Low-Latency Communication (URLLC), a new generic service in 5G wireless systems that puts extreme requirements on reliability. It is therefore interesting to investigate how random access with diversity transmissions of replicas can support very high reliability levels. The study treats two different models, one with destructive collisions and other where the collided slots are still used through a combining process to contribute to the overall SINR. For these two models, we compared two different types of repetition coding, random and deterministic, respectively. The deterministic codes, designed according to a Steiner system, lead to a significantly superior performance when the arrival rate of the intermittent URLLC transmissions is low. 

An interesting direction for future work can be identified in the model with non-destructive collisions. Namely, the current combining algorithm does not take into account that the interference is created from signals that are also packet replicas, just from different users. This fact can be used to devise a multi-user decoding that is capable to cancel the interference from different users, similar to the mechanisms applied in coded random access.


\section*{Acknowledgement}

This work was funded in part by the Academy of Finland (grant 299916) and EIT ICT (HII:ACTIVE) and by the European Research Council (ERC), Consolidator Grant Nr. 648382 (WILLOW), under the European Union Horizon 2020 research and innovation program.



\begin{thebibliography}{00}

\bibitem{5g}
P. Popovski, K. F. Trillingsgaard, O. Simeone and G. Durisi, \emph{``5G Wireless Network Slicing for eMBB, URLLC, and mMTC: A Communication-Theoretic View''}, \emph{{IEEE} Access}, vol. 6, pp. 55765--55779, Aug. 2018 

\bibitem{5gusecase}
3GPP, TS 22.261, 
\emph{``Service requirements for the 5G system; Stage 1,''}
V16.6.0, Dec. 2018

\bibitem{paolini2015}
E. Paolini, C. Stefanovic, G. Liva and P. Popovski, \emph{``Coded random access: applying codes on graphs to design random access protocols''}, \emph{{IEEE} Comm. Mag.}, vol. 53, no. 6, pp. 144--150, June 2015.

\bibitem{choi2017}
J. Choi, ``Throughput Analysis for Coded Multichannel ALOHA Random Access,'' \emph{{IEEE} Comm. Lett.}, vol. 21, no. 8, pp. 1803--1806, Aug. 2017.

\bibitem{singh2018}
B. Singh, O. Tirkkonen, Z. Li and M. A. Uusitalo, ``Contention-Based Access for Ultra-Reliable Low Latency Uplink Transmissions'', \emph{{IEEE} Wireless Comm. Lett.}, vol. 7, no. 2, pp. 182--185, Apr. 2018.

\bibitem{IJWIN}
{C.~Boyd, R.~Vehkalahti and O.~Tirkkonen}, ``{Interference Cancelling Codes
	for Ultra-Reliable Random Access},'' \emph{Int. J. Wireless Inf. Networks}, vol. 25, pp. 422--433, Dec. 2018.	
	
\bibitem{globecomm} 
C. Boyd, R. Vehkalahti and O. Tirkkonen, ``Grant-Free Access in URLLC with Combinatorial
Codes and Interference Cancellation'', in \emph{Proc. {IEEE} Global Comm. Conf.}, Dec. 2018, pp. 1--5.

\bibitem{kotaba2018}
 R. Kotaba, C. N. Manch\'{o}n, T. Balercia and P. Popovski , ``Uplink Transmissions in URLLC Systems With Shared Diversity Resources'', \emph{{IEEE} Wireless Comm. Lett.}, vol. 7, no. 4, pp.  590--593, Aug. 2018.

\bibitem{Paolini}
E.~Paolini, G.~Liva, and A.~G. i~Amat, ``{A Structured Irregular Repetition
	Slotted ALOHA Scheme with Low Error Floors},'' in \emph{Proc. Int. Conf. on
	Comm.}, May 2017, pp. 1--6.

\bibitem{peeters2009}
G. T. Peeters, R. Bocklandt and B. V. Houdt, ``Multiple Access Algorithms Without Feedback Using Combinatorial Designs'', \emph{{IEEE} Trans. on Comm.}, vol. 57, no. 9, pp. 2724--2733, Sep. 2009.

\bibitem{lajolla}
La Jolla Covering Repository. Steiner Systems. [Online]. Available: http://dmgordon.org/cover/

\bibitem{galinina2017}
O. Galinina, A. Turlikov, S. Andreev and Y. Koucheryavy, ``Multi-channel random access with replications'', in  \emph{Proc. IEEE Int. Sym. on Inf. Theory}, Jun. 2017, pp. 2538--2542.




\end{thebibliography}
\end{document}